\documentclass[oneside,twocolumn,superscriptaddress,eqsecnum,amsmath,showpacs]{revtex4-1}
\usepackage{bm}
\usepackage{amssymb}
\usepackage{amsmath}
\usepackage{amsfonts}
\usepackage{braket}
\usepackage{epsfig}
\usepackage{graphicx}
\usepackage{natbib}
\usepackage{mathdots}
\usepackage{multirow}

\usepackage{epstopdf}
\usepackage{hyperref}
\usepackage{float}	
\usepackage{bibentry}
\usepackage[caption=false]{subfig}
\newcommand{\ba}{\begin{eqnarray}}
\newcommand{\ea}{\end{eqnarray}}
\newcommand{\bd}{\begin{displaymath}}
\newcommand{\ed}{\end{displaymath}}
\renewcommand{\v}[1]{{\bf #1}}
\newcommand{\bpm}{\begin{pmatrix}}
\newcommand{\epm}{\end{pmatrix}}
\newcommand{\nn}{\nonumber \\}
\graphicspath{{figures/}}

\begin{document}
\title{Adiabatic Green's function technique and the transient behavior \\
in time-dependent fermion-boson coupled models}
\author{Yun-Tak Oh}
\affiliation{Department of Physics, Sungkyunkwan University, Suwon
  16419, Korea}
\author{Yoichi Higashi}
\affiliation{Department of Physics, Sungkyunkwan University, Suwon
  16419, Korea}
\author{Ching-Kit Chan}
\affiliation{Department of Physics, Massachusetts Institute of Technology, Cambridge, Massachusetts 02139, USA}
\author{Jung Hoon Han}
\email[Electronic address:$~~$]{hanjh@skku.edu}
\affiliation{Department of Physics, Sungkyunkwan University, Suwon 16419, Korea}
\date{\today}
\begin{abstract}
The Lang-Firsov Hamiltonian, a well-known solvable model of interacting fermion-boson system with sideband features in the fermion spectral weight, is generalized to have the time-dependent fermion-boson coupling constant. We show how to derive the two-time Green's function for the time-dependent problem in the adiabatic limit, defined as the slow temporal variation of the coupling over the characteristic oscillator period. The idea we use in deriving the Green's function is akin to the use of instantaneous basis states in solving the adiabatic evolution problem in quantum mechanics. With such ``adiabatic Green's function" at hand we analyze the transient behavior of the spectral weight as the coupling is gradually tuned to zero. Time-dependent generalization of a related model, the spin-boson Hamiltonian, is analyzed in the same way. In both cases the sidebands arising from the fermion-boson coupling can be seen to gradually lose their spectral weights over time. Connections of our solution to the two-dimensional Dirac electrons coupled to quantized photons are discussed.
\end{abstract}
\maketitle

\section{Introduction}
Time-dependent quantum-mechanical phenomena have interested scientists since the inception of quantum mechanics.
A rare example of an exactly solvable time-dependent problem was discovered as early as 1932, known as the Landau-Majorana-Zener problem~ \cite{landau,majorana,zener}. A particular class of time-dependent problems in which the Hamiltonian is periodic in time, $H(t+T)=H(t)$, can be treated in the Floquet framework~\cite{shirley,sambe}. Generalizations of the Floquet theory that include the coupling to the dissipative reservoir and the quench of the periodic drive have been studied extensively in the past\,\cite{jauho1994time,kohler1997floquet-markovian,grifoni1998driven-quantum-tunneling,Ratner2006inelastic,ratner2007molecular,schmidt2009transient,martin-rodero2010nonequilibrium,schmidt2011charge,yeyati2013long,lei2013full,rabani2014nonequilibrium,dehghani2014dissipative-Floquet,martin-rodero2015local,souto2015transient-dynamics}.

The other limit in which the time-dependent problem becomes tractable is when the temporal variation is slow, or ``adiabatic". A general strategy for treating the adiabatic evolution of the quantum system was laid out by Berry~\cite{berry}. The basic idea there was to expand the quasi-exact eigenstate in the instantaneous basis $|\phi_n (t)\rangle$, defined by the eigenvalue problem
\begin{equation} H(t) |\phi_n (t)\rangle = E_n (t) |\phi_n (t)\rangle  \end{equation}
for each time slice $t$.
It is implicit in carrying out Berry's program that one has the solutions of the instantaneous Hamiltonian $H(t)$ at hand.
Berry's idea is most often applied to the single-particle evolution under a parametrically slow external drive,
but the idea itself is general enough to apply to an arbitrary many-body problem,
provided a well-defined gap separates the ground state from the first excited state at all times.
For many-body problems it is often more useful to work with the Green's function containing information for all energies,
instead of the wave function that addresses the ground state property only.
We show how to derive the non-equilibrium Green's function in the adiabatic limit, for a simple time-dependent many-body model.
Explicitly, we work with the time-dependent generalization of the exactly solvable Lang-Firsov (LF) model~\cite{lang-firsov} and a related, spin-boson (SB) model~\cite{Lee_pre,hur2009,hur2015}.
Due to the time dependence of the Hamiltonian the two-time Green's function becomes dependent on the two times separately.
Most often, calculation of the non-equilibrium two-time Green's function is done by the Keldysh technique~\cite{intro_Keldysh,jauho2006introduction}.
We show, in the adiabatic limit of the time-dependent Lang-Firsov model, how to obtain the two-time Green's function without the reliance on the Keldysh method.

Stripped down to its bare minimum, the LF model contains a single fermionic level coupled to a single harmonic oscillator of frequency $\omega_0$. The exact single particle Green's function, obtained through a canonical transformation method, shows in its imaginary part a series of delta function peaks spaced at intervals of $\hbar \omega_0$~\cite{mahanbook}.
Each $n$-th delta function represents a fermionic level dressed by $n$ bosons. When the fermion-boson coupling is turned off, the series of delta functions will reduce to a single peak at the fermion energy. How the evolution from multiple peaks to a single peak takes place, as the coupling is gradually turned to zero, is the question we are going to address quantitatively with the adiabatic Green's function method.

We begin by making a brief discussion of the two-dimensional Dirac electrons coupled to quantized photon fields in Sec. \ref{sec:Dirac}. Although this is not the main focus of our research, it nevertheless helps set the stage for the work done in subsequent sections and lay out the motivation for the problem we choose to solve. In Sec. \ref{sec:LF Hamiltonian} we introduce a time-dependent variant of the LF Hamiltonian and outline how to derive the Green's function for it.  Complex details of the derivation can be found in the Appendix \ref{appendix a}. Recent developments in pump-probe technology have made it possible to observe real-time dynamics of the band electrons under the influence of the intense pump laser. The time-resolved photoemission spectroscopy can be calculated with the lesser Green's function for the system~\cite{freericks}.
We calculate the photo-current for the time-dependent LF model based on our calculation of the lesser Green's function in Sec. \ref{sec:numerics for LF model}, with emphasis on how the sidebands decay over time as the fermion-boson coupling is gradually turned off.
In Sec. \ref{sec:T-dependent SB model}, we solve the time-dependent version of the spin-boson model.
It is our hope that the technology developed in this paper can be further generalized to solve the problem of sideband decay in real materials such as the surface of topological insulators~\cite{gedik-science,mahmood2016selective}.
A summary and outlook is given in Sec. \ref{sec:discussion}.

\section{Dirac electrons coupled to quantized photons}
\label{sec:Dirac}

Although our goal is a simple one - finding solutions to the adiabatic generalization of exactly solvable models - the work we did here may have a non-trivial implication to a much more physical situation of current interest. This is the problem of  two-dimensional (2D) Dirac electrons coupled to the intense laser as studied in several papers in recent years \cite{oka-aoki,galitskii-refael,galitskii-gedik,gedik-science,rigol2015,Gedik-naturephy}.
The Hamiltonian for 2D Dirac electrons coupled to the laser is given by
\begin{align}
H(t) &= \int d \v r ~\psi^\dag(\v r) \left[v \left( -i \bm \nabla + e \v A(\v r , t) \right) \times  {\bm \sigma} \cdot \hat{\v y} \right]\psi(\v r) ,
\label{eq:2d dirac}
\end{align}
where $\bm \sigma$ are the Pauli matrices, $\hat{\v y}$ is normal to the 2D surface, and $\psi(\v r)  =  (u(\v r) , d(\v r)) ^{\rm T}$ are the real-space versions of the spin-up $u_\v k$ and spin-down $d_\v k$ operators. We have set $\hbar=1$.
The vector potential for the laser light is written in the following second-quantized form:
\begin{equation}
{\v A}({\v r},t) = \sum_{\v p} \sqrt{\frac{1}{2 \epsilon_0 \omega_\v p V}} \left(\v \epsilon_{\v p} a_{\v p } e^{i \v p \cdot \v r-i \omega_\v p t} + {\rm h.c.} \right).
\end{equation}
Here, $\epsilon_0$ is the dielectric constant, $\omega_\v p$ is the frequency of laser at momentum $\v p$, and $V$ is the volume of the box.  We can choose a monochromatic frequency for the laser $\omega_\v p = \omega_0$, and the perpendicular direction of incidence for which $\v \epsilon_\v p \cdot \hat{\v y} = 0$. Keeping the ${\v p} ={\v 0}$ component of the vector potential only gives
\ba
H(t) &  \approx & \sum_\v k \psi_\v k^\dag \left[ v (\v k - e  \v A_{\v p = \v 0}(t) )\times
\bm \sigma \cdot \hat{\v y} \right] \psi_\v k , \nn
{\v A}_{\v 0}(t) & = &  {g \omega_0 \over ev} \left({\v \epsilon}_{\v 0} a_{\v 0} e^{-i \omega_0 t} + {\v \epsilon}^*_{\v 0} a^\dag_{\v 0} e^{i \omega_0 t} \right), \label{eq:2d dirac_2}
\ea
where $g =  e v/\sqrt{2 \epsilon_0  \omega_0^3 V}$.
For the linear polarization of the incident laser we can choose $\v \epsilon_\v 0 =  (1, 0, 0)$, and the Hamiltonian becomes
\begin{align}
H  =  \sum_\v k \psi_\v k^\dag \Bigl[& \left(- v k_x + g \omega_0 ( a_\v 0 + a_\v 0^\dag ) \right)\sigma_z \nonumber + v k_z \sigma_x  \Bigr]  \psi_\v k \nn
&+ \omega_0 a^\dag a,
\label{SB-physical}
\end{align}
The photon Hamiltonian is given as a harmonic oscillator, which compensates the dropped time-dependence $e^{\pm i \omega_0 t}$.

Despite the simple appearance of Eq. (\ref{SB-physical}), there is difficulty in solving this problem due to the fact that electrons with different momenta $\v k$ are all coupled to the single photon mode $a_\v 0$ and thereby coupled with each other, somewhat like the single-impurity Kondo problem. However, if we consider a situation in which each electron at momentum $\v k$ couples to a photon mode independently, the problem becomes

\begin{align}
H = \sum_\v k \psi_\v k^\dag \Bigl[ & \left(\frac{\varepsilon_\v k}{2} + g \omega_0 (a_\v k +a_\v k^\dag ) \right)\sigma_z + \frac{\Delta_\v k}{2} \sigma_x  \Bigr] \psi_\v k \nn
&+ \omega_0 \sum_{\v k} a^\dag_{\v k} a_{\v k} .
\label{eq:2d dirac_4}
\end{align}
Here, $\varepsilon_\v k  = -2 v k_x$ and $\Delta_\v k = 2 v k_z$.
One can see that each momentum sector of this Hamiltonian is a realization of the well-known spin-boson model,
widely used in theories of quantum optics, quantum dissipation, quantum computation, and circuit quantum electrodynamics~\cite{Lee_pre, hur2009,hur2015}.
In the limit of $\Delta_\v k \rightarrow 0$, i.e. $k_x \rightarrow 0$, the SB model reduces to the LF model.
In this regard, one can connect the 2D Dirac system coupled with the quantized laser field to the LF model.
The Floquet theory does not work for the 2D Dirac model coupled to the quantized radiation field. The quenching of the laser pulse, which is a critical aspect in the time-resolved ARPES experiments, can be mimicked by the time dependence of the coupling $g=g(t)$. Although this independent photon coupling is a crude approximation, we believe that our analytical treatment of the LF and SB problem can serve as the first step towards the challenging goal of solving the 2D Dirac problem interacting with quantized light field.

\section{Time-dependent Lang-Firsov Hamiltonian}
\label{sec:LF Hamiltonian}

\subsection{The Model}
The Lang-Firsov Hamiltonian
\begin{equation}
H = \varepsilon c ^\dag c + \omega_0 a^\dag a + g \omega_0c^\dag c \left( a + a^\dag \right)
\label{eq:Lang_Firsov_Hamiltonian}
\end{equation}
expresses the coupling of a fermionic level of energy $\varepsilon$ interacting with the harmonic oscillator mode of frequency $\omega_0$.
It is diagonalized by the unitary operator ${\cal U}$:
\begin{align}
{\cal U} &= e^{g c^\dag c( a- a^\dag)}, \nn
\bar{H} &= {\cal U}^\dag H {\cal U} = \bar{\varepsilon}c^\dag c + \omega_0 a^\dag a,
\label{eq:Time Independent LFH}
\end{align}
with the renormalized energy $\bar{\varepsilon} = \varepsilon - g^2 \omega_0 $.
The unitary operator ${\cal U}$ transforms the boson and fermion operators
\begin{align}
{\cal U}^\dag a {\cal U} & =  a -  gc^\dag c,\nn
{\cal U}^\dag c {\cal U} & =  cX,
\label{eq:unitary transforms in lang-firsov}
\end{align}
where one can recognize $X=e^{g (  a-a^\dag) }$ as the coherent state operator,
\begin{align}
X^\dag a  X &= a - g,\nn
X^\dag |\alpha\rangle &= e^{ -\frac{g}{ 2} ( \alpha- \alpha^* ) }  \left| \alpha +g \right\rangle.
\label{eq:coherent shift transforms in lang-firsov}
\end{align}
The factor $e^{ -(g/2)( \alpha- \alpha^* ) }$ is a pure phase and we have introduced the coherent state $|\alpha\rangle$: $a|\alpha\rangle = \alpha |\alpha\rangle$.

The fermion Green's function for the Lang-Firsov model can be obtained exactly thanks to the existence of a unitary operator ${\cal U}$. For instance, the greater Green's function
\begin{equation}
G^>(t,t') = - i  {\rm Tr}[ c(t)c^\dag(t') \rho  ],
\label{eq:Green-one}
\end{equation}
where $\rho$ is the density matrix giving the initial preparation of the fermion-boson state at time $t_0$,
and $c(t) = e^{iH (t-t_0) } c e^{-i H (t-t_0 )}$ is the Heisenberg operator, can be obtained exactly for the initial density matrix
\begin{equation} \rho = |\alpha \rangle \langle \alpha |. \end{equation}
We set empty fermion state because the occupied fermion state gives zero to Eq .(\ref{eq:Green-one}).
A straightforward calculation finds
\begin{align}
G^>(t,t')
=&  -i e^{ -i \bar{\varepsilon}(t-t')} e^{ g^2 \left(e^{-i \omega_0 (t-t') } - 1 \right) } \nn
&  \times e^{( \alpha -\alpha^* ) ( g (t) - g (t') )  },
\label{eq:calculation of Green's function for lang-firsov}
\end{align}
where $g(t) = g e^{i\omega_0 (t -t_0) }$.
When $\alpha = 0$, it reduces to the well-known form
\begin{align}
G^>(t-t') &= e^{-i \bar{\varepsilon}(t-t') -g^2} \sum_{n=0}^\infty g^{2n} \frac{e ^{-i n \omega_0(t-t')} }{ n! } ,
\end{align}
that gives a series of delta-function peaks of weights $g^{2n} /n!$ for the $n$-th sideband.

We now generalize the Lang-Firsov model to include the explicit time dependence in the coupling constant, $g \rightarrow g(t)$:
\begin{equation}
H(t) = \varepsilon c^\dag c + \omega_0 a^\dag a + g(t)  \omega_0  c^\dag c  \bigl(  a+   a^\dag \bigr).
\label{eq:time-dependent-LF}
\end{equation}
This $g(t)$ is not the same factor $g(t)$ appearing in Eq. (\ref{eq:calculation of Green's function for lang-firsov}). Rather, it is a genuine time-dependent fermion-boson coupling $g(t)$ that, by assumption, varies slowly on the time scale of the oscillator $\tau_0 = 2\pi / \omega_0 $,
\begin{equation}
|g' (t) | \tau_0  \ll |g(t)| ,
\label{eq:adiabatic assumption}\end{equation}
where $g'(t)$ is the temporal derivative of $g(t)$.
The Green's function (\ref{eq:Green-one}) for the time-dependent LF model is
\begin{equation}
G^>(t,t') = - i \langle \alpha | U (t_0 ,t)cU (t,t')c^\dag U(t', t_0 ) |\alpha\rangle.
\label{eq:Lang_Firsov_time dependent Green's function}
\end{equation}
The initial time $t_0$ is usually set to the distant past $t_0 \rightarrow -\infty$.
The evolution operator $U (t,t')$, not to be confused with the unitary operator ${\cal U}$ in Eq. (\ref{eq:unitary transforms in lang-firsov}) and (\ref{eq:coherent shift transforms in lang-firsov}), is given by the time-ordered product,
\begin{align}
U(t,t') = T\left[ \exp \left( -i \int^t_{t'} H(t_1) dt_1  \right) \right],
\label{eq:U-propagator}
\end{align}
with the time-dependent LF Hamiltonian (\ref{eq:time-dependent-LF}) in the exponent.
An exact evaluation of the double-time Green's function (\ref{eq:Lang_Firsov_time dependent Green's function}) rests on the exact calculation of the propagator $U(t,t')$, which is not possible in general. On the other hand, the only time dependence in $H(t)$ is through the coupling function $g(t)$, which makes $U(t,t')$ quite close to the propagator $e^{-i(t-t')H}$ of the time-independent Hamiltonian, at least for sufficiently slowly varying $g(t)$ and over a sufficiently small time interval $t-t'$. It suggests that there may be a scheme to systematically expand the propagator $U(t,t')$ in powers of the derivative $g'(t)$. Indeed we have found such a scheme as outlined below.

\subsection{Derivation of the adiabatic Green's function}\label{sec:Evaluation of the GFn}
One can re-write $U(t,t')$ in Eq. (\ref{eq:U-propagator}) as a product over discrete time slices in the spirit of Feynman,
\begin{equation}
U(t,t') = e^{-i \Delta t H(t) } \cdots e^{-i \Delta t H(t_{i}) }\cdots e^{-i \Delta t H(t') },
\label{eq:Lang_Firsov_expansion of time ordering operator}
\end{equation}
and note that any given $e^{-i \Delta t H(t_i )}$ can be diagonalized exactly by the time-dependent unitary operator,
${\cal U}(t_i )$:
\begin{align}
\bar{H}(t_i ) &={\cal U}^\dag (t_i ) H(t_i ) {\cal U}(t_i )=  \bar{\varepsilon}(t_i ) c^\dag c + \omega_0 a ^\dag a,\nn
{\cal U}(t_i ) &= e^{g(t_i)  c^\dag c (a- a^\dag)  },\nn
\bar{\varepsilon}(t_i ) &= \varepsilon - g(t_i )^2 \omega_0.
\label{eq:modulated epsilon}
\end{align}
The replacement
\begin{equation} e^{-i \Delta t H(t_i )} \rightarrow {\cal U}(t_i ) e^{-i \Delta t \bar{H}(t_i ) } {\cal U}^\dag (t_i ) \end{equation}
in Eq. (\ref{eq:Lang_Firsov_expansion of time ordering operator}) gives another expression of the propagator
\begin{align} U(t,t')=
&\Bigl( {\cal U}(t)  e^{-i \Delta t \bar{H}(t) } {\cal U}^\dag(t) \Bigr)  \nn
& \cdots \Bigl( {\cal U}(t_{i})e^{-i \Delta t \bar{H}(t_{i}) }{\cal U}^\dag(t_{i}) \Bigr) \nn
& ~~~ \cdots \Bigl( {\cal U}(t') e^{-i \Delta t \bar{H}(t') } {\cal U}^\dag (t' ) \Bigr) .
\label{eq:LF_unitary tranf for propagator}
\end{align}
The essential idea here is the use of ``instantaneous unitary operator" ${\cal U}(t_i)$ with which to diagonalize the evolution operator $e^{-i \Delta t H(t_i )}$ locally in time.

Another way to organize the product (\ref{eq:LF_unitary tranf for propagator}) is
\begin{equation}
\cdots e^{-i \Delta t \bar{H}(t_{i+1} ) } \Bigl[  {\cal U}^\dag (t_{i+1}){\cal U}(t_i)  \Bigr]  e^{-i \Delta t \bar{H}(t_{i} ) } \cdots .
\end{equation}
Due to the fact that unitary operators ${\cal U}(t_i)$ at different time slices do not commute, there is a factor ${\cal U}^\dag (t_{i+1}){\cal U}(t_i)$ sandwiched between a pair of adjacent exponentials $e^{-i \Delta t \bar{H}(t_{i+1} ) }$ and $e^{-i \Delta t \bar{H}(t_{i} ) }$ in the product (\ref{eq:LF_unitary tranf for propagator}). Since the time difference $t_{i+1}-t_i = \Delta t$ is by assumption very small, one can ignore the small non-commuting factor of order $(\Delta t)^2$ and combine the product ${\cal U}^\dag (t_{i+1}){\cal U}(t_i)$ as \cite{werner2013phonon}
\begin{equation}
{\cal U}^\dag (t_{i+1}) {\cal U}(t_i)
\approx e^{ -  g'(t_i ) \Delta t \; c^\dag c  \left( a-  a^\dag \right) },
\end{equation}
In other words, the exact propagator $U(t,t')$ is obtained from path-ordered exponential of the new effective Hamiltonian
\begin{align}
I(t) &=\bar{H} (t) - i g'(t) c^\dag c \left(  a-  a^\dag\right),\nn
\bar{H}(t) &=  \bar{\varepsilon} (t)  c^\dag c + \omega_0 a ^\dag a,
\label{eq:I}
\end{align}
as
\begin{align}
U(t,t') &={\cal U}(t) \bar{U}(t,t') {\cal U}^\dag(t'). \nn
\bar{U}(t,t')&=T\left[ \exp \left( -i \int^t_{t'} I (t_1) dt_1  \right) \right] . \label{eq:U-diagonal}
\end{align}
The new Hamiltonian $I(t)$ contains the first derivative of the coupling, $g'(t)$, not $g(t)$ itself, and much more conductive to perturbative treatment in powers of the small function $g'(t)$. Another way to view $I(t)$ is as a time-dependent unitary rotation
\begin{equation}
I(t) =  {\cal U}^\dag(t) H(t) {\cal U}(t) - i  {\cal U}^\dag(t) \partial_t {\cal U}(t)
\end{equation}
which yields the same expression as Eq. (\ref{eq:I}). Note that Eq. (\ref{eq:U-diagonal}) is still an exact writing of the propagator.

The next stage of evaluation involves some perturbative scheme, under the adiabaticity assumption.
We have developed the interaction picture scheme to write down the propagator as a power series in $g'(t)$. Details are involved and can be found in the Appendix \ref{appendix a}.
Here, we just quote the zeroth-order result for the Green's function.
\begin{widetext}
\begin{align}
&G^{>,(0)} (t,t') =  -i e^{-i \int^t_{t'} dt_1 [  \bar{\varepsilon}(t_1) -   g'(t_1)^2/ \omega_0] } \langle \alpha |  X(t) cc^\dag X^\dag(t') |\alpha\rangle, \nn
& \langle \alpha |  X(t) cc^\dag X^\dag(t') |\alpha\rangle  =  \exp\Bigl[ g(t) g(t') e^{-i \omega_0 (t-t')} -\frac{1}{2}\Bigl( g(t)^2   + g(t')^2    \Bigr)\Bigr]\nn
& \times \exp\Bigl[ \alpha e^{i \omega_0 t_0} \left( g(t)  e^{-i \omega_0 t}-g(t') e^{- i \omega_0 t'} \right)\Bigr] \times \exp\Bigl[-\alpha ^* e^{-i \omega_0 t_0 } \left(g(t )e^{i \omega_0 t}-g(t' )e^{i \omega_0 t'} \right) \Bigr] .
\end{align}
\end{widetext}
We label it the adiabatic Green's function for an obvious reason. While it is difficult to compare the validity of this Green's function against an exact one for general $g(t)$, our calculation in the following section confirms that corrections up to the second order make negligible difference to the zeroth-order one given above. Although a vast amount of literature was devoted to the study of time-dependent and transient dynamics in quantum models, we believe this is the first time that the Green's function valid in the adiabatic limit is explicitly written down.

\section{Transient behavior of the Green's function} \label{sec:numerics for LF model}

According to Ref. \onlinecite{freericks}, the time-resolved photoemission spectroscopy (TR-PES) intensity at the binding energy $\omega$, $P(t_p ,\omega)$, is obtained from
the formula
\begin{align}
P(t_p,\omega) & \approx -i \int_{-\infty}^\infty dt_2 \int_{-\infty}^\infty  dt_1 s(t_1- t_p) \, s(t_2 - t_p)  \nn
&  ~~~~~ \times e^{i \omega (t_1-t_2) } G^< (t_1,t_2) .
\label{eq:pump-probe intensity integration}
\end{align}
The probe pulse shape function $s(t-t_p )$ is determined by the specific experimental setup. We choose the step-function profile
\ba  s(t-t_p) = \theta(t-t_p) - \theta(t-\sigma_{\rm pr}-t_p) \ea
that corresponds to the probe pulse duration $t_p < t < t_p +\sigma_{\rm pr}$. $P(t_p,\omega)$ records the total accumulated photo-current over the pulse duration $\sigma_{\rm pr}$ which started at time $t_p$. Reference \cite{sentef_prx} showed that  the resolution of TR-PES $\sigma_{\rm res}$ is proportional to the inverse of temporal width of probe pulse; $ \sigma_{\rm pr} \sim 1  / \sigma_{\rm res}$. Since we want to make $\sigma_{\rm res}  \ll \omega_0 $, we set $\sigma_{\rm pr} =  10 \tau_0$, where $\tau_0 =  2 \pi / \omega_0$. The non-equilibrium system itself is prepared at time $t_0$ which is set at the far past.
Throughout the time evolution $t_0< t< t_p + \sigma_{\rm pr}$ the system is governed by the time-dependent LF Hamiltonian $H(t)$.

The lesser Green's function $G^< (t,t')$ in the intensity formula
\begin{equation}
G^< (t,t') = i {\rm Tr} [ c^\dag(t') c(t)  \rho ]
\end{equation}
is different from the one analyzed in the previous section, can be solved with the same technology. There is a certain degree of freedom in choosing the initial state $|\psi\rangle$ and the initial density matrix $\rho = |\psi\rangle \langle \psi |$. Our choice for $|\psi\rangle$ is a product of the boson coherent state and a one-electron state, hybridized by the unitary operator ${\cal U}$:
\begin{equation}
|\psi\rangle = {\cal U} \Bigl( c^\dag |\alpha\rangle \Bigr) = e^{(g/2) \left(  \alpha - \alpha^* \right)} c^\dag |\alpha- g\rangle.
\end{equation}
Using $\rho = |\psi\rangle \langle \psi |$,
\begin{equation}
G^{<} (t,t') =   i \langle \alpha-g| c c^\dag(t') c(t) c^\dag |\alpha - g \rangle.
\end{equation}
Unlike the greater Green's function case, we set occupied fermion state since empty fermion state gives zero to the lesser Green's function.
Evaluating the lesser Green's function yields
\begin{widetext}
\begin{align}
&G^{<,(0)} (t,t')  =  i \exp\left[-i \int^t_{t'}dt_1 \left(\bar{\varepsilon}(t_1) - \frac{g'(t_1)^2}{ \omega_0} \right)\right]\times \exp \left[g(t)\,g(t') e^{i \omega_0(t-t')}-\frac{1 }{ 2} \left(g(t)^2 +  g(t')^2 \right)  \right] \nn
& ~~~~~ \times \exp\left[ \alpha e^{ i \omega_0 t_0} \left( g(t) e^{-i \omega_0 t} -g(t') e^{-i \omega_0 t'}  \right)\right] \times \exp\left[-\alpha^* e^{-i \omega_0 t_0} \left(g(t) e^{i \omega_0 t} -g(t') e^{i \omega_0 t'}  \right) \right].
\label{eq:zeroth ordered lesser Green's function}
\end{align}
\end{widetext}
We also obtained the first and second corrections for lesser Green's functions, $G^{<,(1)}(t,t')$ and $G^{<,(2)}(t,t')$, as reproduced in the Appendix \ref{appendix b}. The total lesser Green's function up to second order in the derivative $g'(t)$ is the sum,
\begin{equation}
G^<(t,t') = G^{<,(0)}(t,t')+G^{<,(1)}(t,t')+G^{<,(2)}(t,t').
\label{eq:total lesser Green's function}
\end{equation}
An explicit numerical evaluation finds negligible contributions to the photo-current from higher-order Green's functions $G^{<, (1)}$ and $G^{<,(2)}$, making the zeroth-order Green's function we derived in Eq. (\ref{eq:zeroth ordered lesser Green's function}) essentially exact for the time-varying coupling $g(t)$. The conditions for their validity are that the typical variation in $g(t)$ occurs over a time scale much longer than the oscillator period, $|g' (t) \tau_0 / g(t)| \ll 1$, and that $g' (t)$ itself varies little over one period $\tau_0$. The second assumption however is natural in light of the first.
Given that the typical pump laser in use today operates at the sub-visible range, $\omega_0 \sim 10^{13}$Hz, this is a rather comfortable assumption to be made.

\begin{figure}[htbp]
\includegraphics[width=0.45\textwidth]{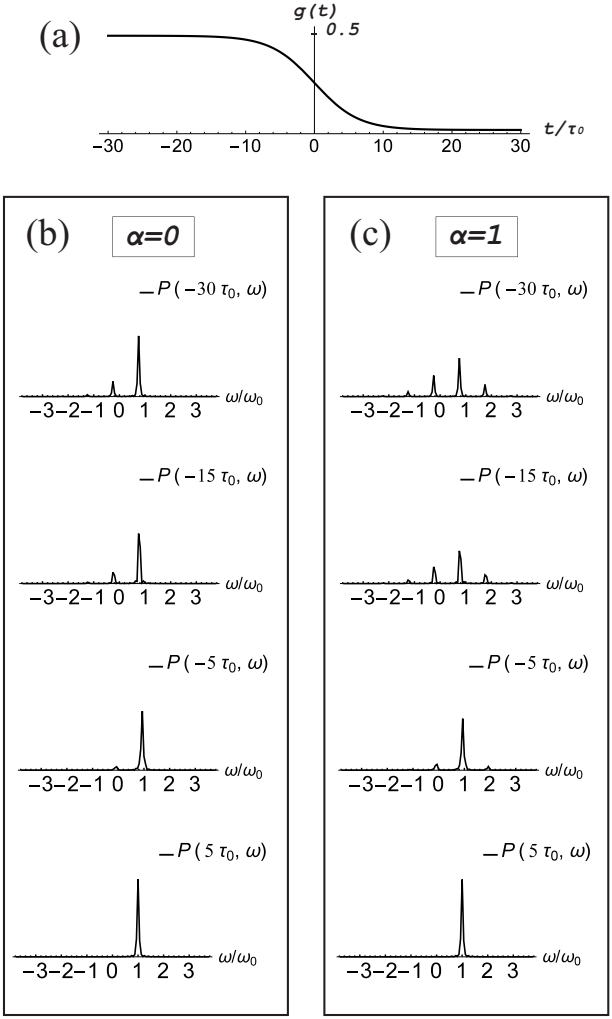}
\caption{Profile of the coupling $g(t)$ over time.
Frequency resolved photo-current intensity $P(t,\omega)$ at various times for the coherent state amplitudes (b) $\alpha=0$ and   (c) $\alpha=1$. Fully developed sideband feature at $\omega=\bar{\varepsilon} + ({\rm integer}) \omega_0$ shown at $t=-30\tau_0$ start to slide over to the higher frequency for times past $t=-15 \tau_0$ where $g'(t)$ also starts to vary.
The bare fermion energy peak vanishes at $t = 5 \tau_0$.}
\label{fig:1}
\end{figure}

How will the photo-current intensity $P(t,\omega)$ evolve over time as the electron-boson coupling $g(t)$ is adiabatically turned off to zero? To explore this, we proceed to the numerical evaluation of $P(t,\omega)$ using the profile
\begin{equation}
g(t) = \frac{g }{ e^{t /T_g} +1 }
\label{eq:LF_g_profile}
\end{equation}
for the coupling function $g(t)$. Here, we set $T_g = 3.2 \tau_0$. The adiabatic condition
\ba  \left| { g' (t) \tau_0 \over g(t) }  \right|  =  {\tau_0 \over T_g } { 1\over e^{t / T_g} +1}  \lesssim {\tau_0 \over 2T_g }   \ll 1  \ea
is fulfilled at all times $|t| \lesssim T_g$.
The observation time $t$ extends from $-45\tau_0$ up to $45\tau_0$ in our calculation.
The initial preparation time $t_0$ is set further back at $t_0 = -130\tau_0$.
For the parameters of the model we choose $g= 0.5$ and $\varepsilon=  \omega_0 $, which gives the renormalized energy $\bar{\varepsilon}=0.75\omega_0$.
The lesser Green's function can be obtained numerically for various choices of the coherent state $\alpha$.
The phase angle in $\alpha$ can be absorbed since it always appears as the product $\alpha e^{i\omega_0 t_0}$ in the Green's function [see Eq. (\ref{eq:pump-probe intensity integration})].

From the calculations, it turns out the two higher-order Green's functions in orders of $g'$ and $(g')^2$ make negligible contributions to the photo-current $P(t, \omega)$ for the $g(t)$ chosen in Eq. (\ref{eq:LF_g_profile}).
For the photocurrent intensity $P^{(i)}(t,\omega)$ with superscript $i=\{1,2\}$ denotes the order of $g'$ and $(g')^2$ contribution, we found that even the maximum of $|\int d \omega P^{(i)}(t,\omega)/ \int d \omega P^{(0)}(t,\omega)|\approx 10^{-7}$, which is negligible, for the entire $t$ in our calculation. We conclude that it suffices to discuss the photo-current obtained from the zeroth-order $G^{<,(0)}$ alone.
In this regard the adiabatic method we developed to obtain the two-time  Green's function for the time-dependent LF Hamiltonian is already {\it exact} at the zeroth order.

Figure \ref{fig:1}  shows the photo-current $P(t,\omega)$ at several times $t$ throughout the adiabatic turn-off of the coupling $g(t)$.
Several sidebands, present at times long before the adiabatic turn-off process began, have their frequencies shifted by  $\sim g^2 \omega_0 = 0.25 \omega_0$ as $g$ diminishes to zero.  Their intensities diminish over time. The main peak at the energy $\bar{\varepsilon}(t)$ also slides in frequency by $g^2 \omega_0 =0.25 \omega_0$ with its intensity growing over time.
Even for $\alpha = 0$ case, we can see that the sidebands emerge at $\omega = \bar{\varepsilon}  - n \omega_0$ ($n > 0$).
These sidebands for $\alpha=0$ are coming from the terms proportional to the $g^2$ in Eq. (\ref{eq:zeroth ordered lesser Green's function}).

It is notable that we obtain the diminishing sidebands feature even without manifestly introducing the dissipation mechanism such as the bosonic bath, explicitly~\cite{dehghani2014dissipative-Floquet}. In an adiabatic evolution of the quantum system such as the expanding potential well, the instantaneous energy of the system smoothly follows the ground state value of the instantaneous Hamiltonian. As the wall expands the energy also diminishes, but this is done without an explicit dissipation mechanism. The same phenomenon is happening in our Green's function treatment of the adiabatic evolution.

\subsection{Transient behavior in the semi-classical limit}
The boson field is treated as a quantized oscillator in our approach to transient dynamics. In this subsection, we ask
what happens if the boson field is treated semi-classically, and the relevant Hilbert space is that of fermions only. The semi-classical limit of the time-dependent LF Hamiltonian is obtained by going to the interaction picture, $a\rightarrow a e^{-i\omega_0 (t-t_0)}$,
\begin{align}
\hat{H}_{\rm LF} (t)
& =  e ^{i \int^t_{t_0} dt_1 H_b} H_{\rm LF} (t) e ^{-i \int^t_{t_0} dt_1 H_b}  -\omega_0 a^\dag a \nn
& = \varepsilon c^\dag c + g(t) \omega_0 c^\dag c\left(a  e^{-i \omega_0 (t-t_0 ) }+ a^\dag e^{i \omega_0 (t-t_0) } \right) ,
\end{align}
and then replacing $a$ by its average $\langle a \rangle =\alpha$, assuming a coherent state of the boson:
\begin{align}
H^{cl.}_{\rm LF} (t)
& = \varepsilon c^\dag c + 2 g(t) \omega_0  \alpha  \cos [\omega_0 (t-t_0 )] c^\dag c .
%
\label{eq:cl_LF S}
\end{align}

The lesser Green's function for the semi-classical, time-dependent LF model is still of the form,
\begin{align}
G^<(t,t') =   i {\rm Tr} \left[\rho(t_0) c^\dag(t')  c(t) \right] .
\end{align}
The Hilbert space is now confined to the two-level fermion states only, and the density matrix $\rho(t_0) = |\psi\rangle \langle \psi|$
consists of the one-fermion state $|\psi\rangle =  c^\dag|0\rangle$. The lesser Green's function for arbitrary coupling $g(t)$ becomes
\begin{align}
&G^<(t,t') \nn
&= i e^{ -i \varepsilon(t-t') - i \int^t_{t'}dt_1 2 g(t_1) \omega_0 \alpha \cos \omega_0 (t_1-t_0) } \nn
& \approx i e^{-i \varepsilon(t-t')} \exp[ \alpha   e^{-i \omega_0  t_0} (g(t)  e^{i \omega_0 t} -g(t') e^{i \omega_0 t'}) ]\nn
&\times \exp[- \alpha e^{i \omega_0  t_0} (g(t)  e^{-i \omega_0 t} -g(t') e^{-i \omega_0 t'}) ].
\label{eq:cl_LF lesser Green's function}
\end{align}
\begin{figure}[htbp]
\includegraphics[width=0.45\textwidth]{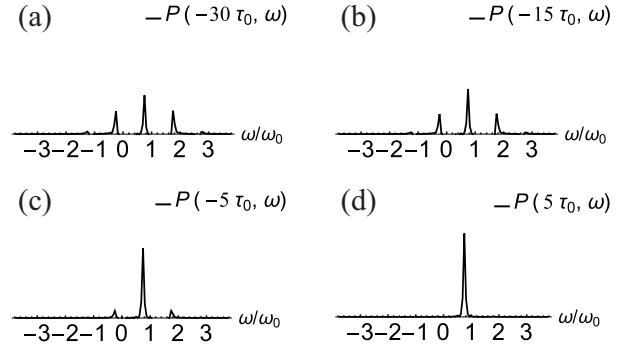}
\caption{ Frequency-resolved photo-current intensity $P(t,\omega)$ at various times for semi-classical LF model with $\alpha=1$ and $\varepsilon = 0.75 \omega_0$.
Fully developed sidebands at $\omega = \varepsilon + ({\rm integer}) \omega_0$ are shown in (a).
For classical case, the intensities of sidebands have bilateral symmetry with respect to the bare electron energy $\varepsilon$.
In (b)-(d), the intensity corresponding to the bare electron energy monochromatically increases as time evolves while other sidebands' intensities monochromatically decrease.  }
\label{fig:3}
\end{figure}
In the last line we have ignored terms proportional to $g'(t)$, as allowed by the adiabatic assumption Eq. (\ref{eq:adiabatic assumption}). One can easily notice that Eq. (\ref{eq:cl_LF lesser Green's function}) can be recovered by erasing the terms proportional to the $g^2$ in the arguments of exponentials in Eq. (\ref{eq:zeroth ordered lesser Green's function}).

From the semi-classical, time-dependent Green's function (\ref{eq:cl_LF lesser Green's function}) we obtain the photo-current shown in Fig. \ref{fig:3}. We have used the identical profile for the coupling function $g(t)$ as in the earlier, quantum-mechanical LF model [Eq. (\ref{eq:LF_g_profile})] with $g=0.5$. We set the other parameters $\alpha =1$ and $\varepsilon = 0.75 \omega_0$.
Again, in the photo-current calculation the probe beam starts at $t_p = - 45 \tau_0$ and observation time end at $t_p = 45 \tau$.
The initial preparation time $t_0$ is set at $t_0 = -130\tau_0$. Initially, as shown in Fig. \ref{fig:3}(a), there are well-developed sideband peaks in the semi-classical photo-current $P(t,\omega)$ as well. As one turns $g(t)$ off the weights at sideband energies diminish and only the weight at the bare energy $\varepsilon$ grows monotonically.

A number of subtle differences exists between semi-classical and quantum calculations of the photo-current profile.
First, since there is no renormalization of the bare electron level $\varepsilon$ in the semi-classical limit, there cannot exist the ``sliding over'' feature of the peaks of photocurrent intensities. Next, the profile $P(t_p ,\omega)$ in the semi-classical calculation remains completely symmetric about $\omega=\varepsilon$ at all times $t_p$ since there are no spontaneous emission of boson in the classical limit. The sidebands of semi-classical calculation are fully due to the terms proportional to $\alpha$ in Eq. (\ref{eq:cl_LF lesser Green's function}). Even with these subtle differences, it is notable that the semi-classical Green's function is recovered in the large $\alpha$ limit of Eq. (\ref{eq:zeroth ordered lesser Green's function}). This fact is consistent with the idea of considering boson field classically in the limit of large number of boson $N = |\alpha|^2$. Since the calculation for the semi-classical calculation is straightforward, the fact that the Green's function in Eq. (\ref{eq:zeroth ordered lesser Green's function}) is recovered by the semi-classical Green's function supports that our method is reasonable.

\subsection{Time-dependent spin-boson model} \label{sec:T-dependent SB model}
Techniques we developed to address the transient phenomena in the Lang-Firsov model with time-dependent coupling can be applied, with a little modification, to another well-known and popular spin-boson (SB) model describing the two-level system interacting with the bosonic field:
\begin{align}
H_{\rm SB} = \frac{\varepsilon}{2} \sigma_z + \frac{\Delta}{2} \sigma_x  + \omega_0 a^\dag a + {g\over 2} \omega_0 \sigma_z\left( a + a^\dag \right).
\end{align}
This model for $\Delta=0$ is none other than the Lang-Firsov Hamiltonian by replacing $\sigma^+ \rightarrow c^\dag$, $\sigma^- \rightarrow c$, and $\sigma^z = 2c^\dag c -1$.
The transition term $(\Delta/2) \sigma_x$ between two energy levels does not have a fermion analogue as it corresponds to single fermion annihilation and creation processes $\sim \Delta ( c^\dag + c)$.

Applying the unitary operator ${\cal U}  =  e^{g \sigma_z (a- a^\dag) }$ gives
\begin{align}
\bar{H}_{\rm SB} & =  {\cal U}^\dag H_{\rm SB} {\cal U}\nn
&= \frac{\varepsilon}{2} \sigma_z -{g^2 \over 4}  \omega_0 + \omega_0 a^\dag a
+ \frac{\Delta}{2} \left( X^\dag \sigma^++ X \sigma^- \right),
\end{align}
where $X= e^{ g (a-a^\dag) }$. The interaction term $(g/2) \omega_0 \sigma_z (a+ a^\dag) $ is gone, but there is a residual interaction of order $\Delta$ in the transformed Hamiltonian $\bar{H}_{\rm SB}$. It turns out to be exceedingly difficult to keep both the time dependence of the coupling $g(t)$ and the residual interaction of order $\Delta$ in calculating the adiabatic Green's function. From now on we drop the $\Delta$ piece in the above and generalize the $\Delta=0$ spin-boson Hamiltonian to the time-dependent one:
\begin{equation}
H_{\rm SB}(t) =  \frac{\varepsilon}{2} \sigma_z  + \omega_0 a^\dag a +  {1\over 2} g(t) \omega_0 \sigma_z \left( a + a^\dag \right).
\end{equation}

We assume that $g(t)$ is a slowly varying function in time and define the lesser Green's function for the SB Hamiltonian as
\begin{align}
G^{<}(t,t') =   i {\rm Tr} \left[\rho (t_0 )  \sigma^+(t') \sigma^-(t)   \right] ,
\label{eq:SB_lesser_Green's_Fucntion}
\end{align}
where $\sigma^\pm =\left( \sigma_x \pm i \sigma_y \right) /2$. Choosing the initial state density matrix $\rho (t_0 ) = |\psi \rangle\langle \psi |$, where $|\psi \rangle =  | \uparrow, \alpha - g \rangle $, calculation of the lesser Green's function proceeds in direct analogy with the one for the LF model. We obtain
\begin{widetext}
\begin{align}
G^{<,(0)}(t,t')
& =  i e^{-i \varepsilon(t-t')} \langle \uparrow, \alpha | \sigma^+ \hat{X}(t') \sigma^-\hat{X}(t)|\uparrow,\alpha\rangle \nn
&= ie^{-i \varepsilon(t-t') } \times\exp\left[g(t) g(t') e^{i \omega_0 (t-t')} -{1\over 2} \left(g(t)^2 + g(t')^2 \right)\right]  \nn
&~~~~\times \exp\left[\alpha e^{i \omega_0 t_0} \left(g(t) e^{-i \omega_0 t} - g(t') e^{-i \omega_0 t'} \right)\right]\times \exp\left[- \alpha^* e^{-i \omega_0 t_0} \left(g(t) e^{i \omega_0 t} - g(t') e^{i \omega_0 t'} \right)\right].
\label{eq:SB_lesser_Green's_Fucntion_0th}
\end{align}
\end{widetext}
One can see this expression is almost identical to the zeroth-order Green's function worked out in Eq.
(\ref{eq:zeroth ordered lesser Green's function}).

\begin{figure}[htbp]
\includegraphics[width=0.45\textwidth]{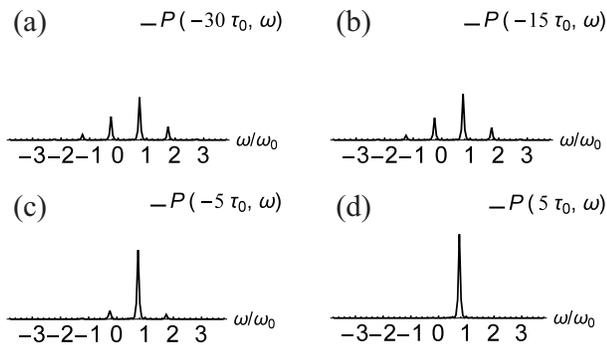}
\caption{Frequency-resolved zeroth-order contribution to the photo-current intensity $P^{(0)}(t,\omega)$ at various times in the time-dependent spin-boson model with $\varepsilon=0.8\omega_0$ and $g=0.5$.
The coherent state amplitude is set to $\alpha=1$.
Fully developed sideband features exist as in the LF model.
The sideband intensity is decreasing monotonically in (b)-(d) after $g(t)$ is turned off, as the intensity for the bare electron energy level increases and eventually saturates. We do not observe any ``sliding over" behavior as in the LF model.
}
\label{fig:sb}
\end{figure}

The transient behavior in the photocurrent intensity $P(t,\omega)$ is shown in Fig. \ref{fig:sb}.
The adiabatic behavior of the photocurrent is showing the smooth decay of the sideband weights over time.
In the SB model the bare energy level $\varepsilon$ does not renormalize; hence we do not observe any ``sliding over" behavior in the adiabatic turn-off process that characterized the transient dynamics of the LF model.


\section{Discussion} \label{sec:discussion}
Understanding the transient dynamics of electron-boson coupled system is of growing theoretical importance as pump-probe type experiments get refined at a rapid pace and begin to demonstrate fascinating phenomena~\cite{gedik-science,hasan,cavalleri}.
In this paper we attempt to give theoretical foundation to addressing the question, ``{\it How do the electronic sidebands die out after the pump laser is turned off?}", by solving in a quasi-exact manner the time-dependent versions of the Lang-Firsov and spin-boson Hamiltonians. Our calculation successfully demonstrates the gradual decay of sidebands after the pump has been decoupled from the electronic system.
Existence of the dissipative environment is not a necessary condition to observe the decay in the adiabatic limit as opposed to the previous study \cite{kohler1997floquet-markovian,grifoni1998driven-quantum-tunneling,schmidt2009transient,rabani2014nonequilibrium,dehghani2014dissipative-Floquet,martin-rodero2015local,souto2015transient-dynamics}.

A key theoretical idea allowing us to obtain the non-equilibrium Green's function is the introduction of ``instantaneous basis" of unitary operators $U(t)$, that diagonalizes the Hamiltonian $H(t)$ exactly through the rotation $U^\dag (t) H(t) U(t)$. The small discrepancy in the unitary operators at infinitesimally separated times $U(t+\Delta t) U^\dag (t)$ can be treated perturbatively provided the time evolution of the parameter $g(t)$ in the Hamiltonian is slow in comparison to the characteristic oscillation frequency $\omega_0$. Readers are alerted to the similarity of our idea to Berry's derivation of the geometric phase, which he accomplished by solving the time-dependent Hamiltonian in the ``instantaneous basis" of eigenstates. Berry's adiabatic solution of the wave function \cite{berry} has an analogue in our approach as the zeroth-order Green's function. Corrections to the adiabatic Green's function can be generated by diligent application of the perturbative quantum field theory technique. In the case of time-dependent Lang-Firsov model those perturbative corrections are proved to have negligible impact on the time-dependent photo-current intensity profile.

A related theoretical investigation of time-dependent electron dynamics in the Holstein model in the context of pump-probe ARPES can be found in Ref. \onlinecite{sentef_prx}.
In their study the time-dependent part is the classical radiation field represented as the Peierls substitution of the momentum.
While many aspects of the relaxation phenomena was discussed in that paper, sideband features and their demise after the quench were not. Also noteworthy is that the method employed in this work is not the proto-typical Keldysh technique.
Our approach is one of directly evaluating the two-time Green's function as accurately as possible, with the results shown in Eq. (\ref{eq:zeroth ordered lesser Green's function}) and (\ref{eq:SB_lesser_Green's_Fucntion_0th}) in essentially exact forms.
A number of works studied the non-equilibrium phenomena in the context of a quantum dot coupled to external leads \cite{grifoni1998driven-quantum-tunneling,Ratner2006inelastic,ratner2007molecular,schmidt2009transient,martin-rodero2010nonequilibrium,schmidt2011charge,yeyati2013long,lei2013full,rabani2014nonequilibrium,martin-rodero2015local,souto2015transient-dynamics}. The dot Hamiltonian is akin to the Lang-Firsov model we study in this paper. We propose that the adiabatic Green's function derived here can be adopted to the more physical situation of a quantum dot under non-equilibrium and time-dependent conditions. As discussed in Sec. \ref{sec:Dirac}, our Green's function approach developed here can also shed some light on the more realistic problem about Dirac fermions coupled to quantized photons.

\acknowledgments{Y.-T.O. was supported by a Global Ph.D. Fellowship Program through the National Research Foundation of Korea (NRF) funded by the Ministry of Education (NRF-2014H1A2A1018320).
JHH acknowledges many insightful discussions on time-dependent phenomena with Patrick Lee and thanks him for hospitality during his sabbatical leave in 2014. Part of the motivation for this work came from discussions with Nuh Gedik. }

\appendix
\section{Completion of Derivation of Green's function $G^>(t,t')$ for time-dependent Lang-Firsov model}
\label{appendix a}
In this appendix we introduce our trick to calculate further of propagator $\bar{U}(t,t')$ in Eq. (\ref{eq:U-diagonal}) to complete the calculation of the Green's function  $G^>(t,t')$.
Since we assumed $g'(t)$ small, we can now treat the interaction as perturbation.
In the interaction picture, the propagator becomes
\begin{equation}
 \bar{U}(t,t')  = e^{-i \int^t_{t_0} dt_1 \bar{H}(t_1)} {\cal S}(t,t') e^{i \int^t_{t_0} dt_1 \bar{H}(t_1)},
\end{equation}
in which each operator $A(t)$ becomes
\begin{equation}
\hat{A} (t) = e^{i \int^t_{t_0} dt_1 \bar{H}(t_1)}  A(t) e^{-i \int^t_{t_0}dt_1 \bar{H}(t_1)},
\end{equation}
There is no need to time-order the exponential $\exp\left[-i \int \bar{H}(t_1) dt_1\right]$ since operators at different times now commute:
$[\bar{H}(t_1 ),\bar{H}(t_2 ) ] = 0$.
One can easily verify necessary properties such as
\begin{equation}
{\cal S}(t,t'') {\cal S} (t'', t') = S(t, t'), ~ {\cal S}(t' , t) = {\cal S}^\dag (t,t'). \nonumber
\end{equation}

It is a simple exercise to derive equations of motion
\begin{align}
&\partial_{t} \hat{a}(t) =  -i \omega_0 \hat{a} (t), ~~ \partial_{t}  \hat{c} (t) = -i \bar{\varepsilon}(t) \hat{c}(t), \nn
&\partial_t {\cal S}(t,t') =  -c^\dag c \, g'(t) \Bigl(  \hat{a}(t) -   \hat{a}^\dag (t) \Bigr){\cal S}(t,t'),
\end{align}
and integrate them to obtain
\begin{align}
\hat{a} (t) & =   e^{-i \omega_0 (t - t_0 )  }  a ,\nn
\hat{c} (t) & =   e^{-i \int_{t_0}^t  \bar{\varepsilon}(t_1 ) dt_1 } c , \nn
\hat{X}(t) &= \exp\left[ g'(t) \left( \hat{a}(t) - \hat{a}^\dag(t)  \right)\right], \nonumber
\end{align}
and
\begin{equation} {\cal S}(t,t') \!=\! T e^{ \int^t_{t'} dt_1  g'(t_1) c^\dag c ( \hat{a}^\dag(t_1) \!-\!   \hat{a}(t_1) )  } ,
\label{eq:Stt}
\end{equation}
in the interaction picture.
The advantage of the interaction picture calculation is that the propagator ${\cal S}(t,t')$, Eq. (\ref{eq:Stt}), depends only on the derivative $g' (t_1)$ - a small quantity by assumption - and can be expanded as a power series. Expanding ${\cal S} (t,t')$ allows evaluation of the Green's function to successively higher orders of accuracy in $g'(t)$.

%
%

Let's write the Green's function, Eq. (\ref{eq:Lang_Firsov_time dependent Green's function}), in the interaction basis.
First we use Eq. (\ref{eq:U-diagonal}) to express $G^>(t,t')$ as
\begin{align}
G^>(t,t')
&=  i \langle \alpha | U (t_0 ,t)cU (t,t')c^\dag U(t', t_0 ) |\alpha\rangle   \nn
&= i \langle \alpha | {\cal U}(t_0 ) \bar{U}(t_0 ,t) {\cal U}^\dag(t)c{\cal U}(t)\bar{U}(t,t') \nn
&\times {\cal U}^\dag(t')c^\dag{\cal U}(t') \bar{U}(t',t_0 ) {\cal U}^\dag(t_0 )|\alpha \rangle \nn
&= i \langle \alpha | \bar{U}(t_0 ,t)   c X(t) \bar{U}(t,t')  c^\dag X^\dag (t' ) \bar{U}(t', t_0 ) |\alpha \rangle .
\label{eq:G-equivalent}
\end{align}
The third line follows from ${\cal U}^\dag (t)c{\cal U}(t) = c X(t)$,
${\cal U}^\dag (t)c^\dag{\cal U}(t) = c^\dag X^\dag(t)$, where $X(t) = e^{  g (t)( a- a^\dag ) }$.
Furthermore we have ${\cal U}^\dag (t_0 ) |\alpha\rangle = |\alpha\rangle$ since $c^\dag c |\alpha\rangle = 0$ due to the absence of fermions in the coherent state $|\alpha\rangle$. Now we go to the interaction picture and re-write $G^>(t,t')$ as
\begin{align}
&G^>(t,t') \nn
&= i \langle \alpha | \bar{U}(t_0 ,t) c X(t)\bar{U}(t,t')  c^\dag X^\dag (t' ) \bar{U}(t', t_0 ) |\alpha \rangle \nn
&=  i \langle \alpha | {\cal S} (t_0 , t) e^{i \int^t_{t_0} \bar{H}(t_1) dt_1 } c X(t) e^{-i \int_{t_0}^t \bar{H}(t_1) dt_1 } {\cal S} (t,t' ) \nn
&~~~\times e^{i \int_{t_0}^{t'} \bar{H}(t_1) dt_1 } c^\dag X^\dag (t' ) e^{-i \int_{t_0}^{t'} \bar{H}(t_1) dt_1 }  {\cal S}(t',t_0) |\alpha\rangle \nn
&= i \langle \alpha | {\cal S} (t_0 , t) \hat{c}(t) \hat{X} (t)  {\cal S}(t,t')  \hat{c}^\dag (t' ) \hat{X} (t' ) {\cal S} (t',t_0 ) |\alpha\rangle \nn
&= i e^{-i \int^t_{t'} dt_1 \bar{\varepsilon}(t_1) } \nn
&~~~\times \langle \alpha |  {\cal S}(t_0 ,t) c  \hat{X}(t) {\cal S}(t,t')c^\dag  \hat{X}^\dag (t' )  {\cal S}(t', t_0 ) |\alpha \rangle.
\end{align}
This is the formally exact expression of the two-time Green's function for time-dependent LF model.
Faithful evaluation of the Green's function becomes possible by systematically expanding $S(t,t')$ as a power series.
By inspection of Eq. (\ref{eq:Stt}) one concludes ${\cal S} (t,t') |\alpha \rangle = |\alpha \rangle$ for the zero-fermion state $|\alpha \rangle$, which means $G^>(t,t')$ further simplifies to
\begin{equation}
G^>(t,t') = i e^{-i \int^t_{t'} dt_1 \bar{\varepsilon}(t_1) } \langle \alpha | c  \hat{X}(t){\cal S}(t,t') c^\dag \hat{X}^\dag (t' )  |\alpha \rangle.
\label{eq:G-simplified2}
\end{equation}
The $\hat{X}(t')$ operator does not change the fermion number, while $c^\dag$ raises it by one.
When the next operator ${\cal S} (t,t')$ acts on the one-fermion state one can replace $c^\dag c$ inside $S(t,t')$ by unity,
so effectively,

\begin{equation}
{\cal S} (t,t') = T\left[ \exp\left( \int^t_{t'} dt_1  g'(t_1 )(\hat{a}^\dag(t_1)  - \hat{a}(t_1))  \right) \right].
\label{eq:final-S}
\end{equation}

The final technical hurdle in the Green's function evaluation is to develop a reliable expansion scheme for ${\cal S}(t,t')$ above. A simple Taylor expansion of the exponent won't work here - although that is how the typical diagrammatic calculation would proceed - due to the time-dependent function $g' (t_1)$ in the integrand.
The first step in this regard is to re-write the propagator ${\cal S} (t,t')$ as a product of integrals over one oscillator period $\tau_0$ each,
\begin{align}
& {\cal S} (t,t') \!=\! \nn
& T \exp \left( - \int^t_{t'+N \tau_0 } dt_1 \,g'(t_1)( \hat{a}(t_1) -  \hat{a}^\dag(t_1)) \right) \times  \nn
&T  \prod_{n=0}^N
\exp \left( -\int^{t' \!+\! n \tau_0 }_{t' \!+\! (n\!-\! 1)\tau_0} dt_1 \,  g'(t_1) (\hat{a}(t_1) \! - \! \hat{a}^\dag(t_1)) \right).
 \label{eq:Lang_Firsov_interaction_propagator_1}
\end{align}
It is understood that the last time slice $[t, t'+N \tau_0]$ covers a fraction of the oscillator period $\tau_0$.
For a particular time region $[t_i , t_i + \tau_0]$ we assume that period $\tau_0$ to be small enough that the time ordering within this temporal region can be ignored. As a result it becomes possible to carry out the integral within each time slice, second part of r.h.s. of Eq. (\ref{eq:Lang_Firsov_interaction_propagator_1}) becomes
\begin{align}
T&\left[ \exp \left(  -\int^{t_i+ \tau_0 }_{t_i} dt_1 g'(t_1)\left(\hat{a}(t_1) -\hat{a}^\dag(t_1) \right) \right) \right] \nn
\approx & 1 -  \int^{t_i+ \tau_0 }_{t_i} dt_1 g'(t_i) \left([ a e^{-i\omega_0 (t_1-t_0) } - a^\dag e^{i \omega_0 (t_1-t_0) }  \right) \nn
        & + \int^{t_i+ \tau_0 }_{t_i} dt_1 \int^{t_1}_{t_i} dt_2
         g'(t_i)^2 \left( a e^{-i\omega_0 (t_1-t_0) } -  a^\dag e^{i \omega_0 (t_1-t_0) }\right) \nn
        & \times  \left([ a e^{-i\omega_0 (t_2-t_0) } - a^\dag e^{i \omega_0 (t_2-t_0) }\right) \nn
\approx & 1 + i \frac{ [ g'(t_i) ]^2 }{ \omega_0 } \tau_0 \approx \exp \left(  i \frac{[ g'(t_i) ]^2 }{ \omega_0 } \tau_0 \right)  .
\label{eq:Lang_Firsov_interaction_propagator_2}
\end{align}
First-order terms in $g'(t)$ vanish from the integration over the full period of the harmonic oscillator, leaving a small, second-order correction from the integration. Since each term in the exponent is small, one can add them and express the result as an integral:
\begin{align}
&T\left[  \prod_{i=0}^N \exp \left( -\int^{t_i+ \tau_0 }_{t_i} dt_1 g'(t_1) ({a}(t_1) - \hat{a}^\dag(t_1)) \right) \right] \nn
&\approx \exp \left(  i\int^{t}_{t'} dt_1\, \frac{ \left[ g'(t_1) \right]^2 }{ \omega_0 }  \right).\label{eq:Lang_Firsov_interaction_propagator_3}
\end{align}

The front exponential part of Eq. (\ref{eq:Lang_Firsov_interaction_propagator_1}) can be analyzed similarly,
\begin{widetext}
\begin{align}
T&\left[ \exp \left( - \int^t_{t'+N \tau_0 } dt_1 g'(t_1)\left( \hat{a}(t_1) - \hat{a}(t_1) \right) \right) \right] \nn
\approx & 1 -\int^t_{t'+N \tau_0 } dt_1 g'(t)  \left( a e^{-i\omega_0 (t_1-t_0) } -   a^\dag e^{i \omega_0 (t_1-t_0) }  \right) \nn
& + \int^t_{t'+N \tau_0} dt_1 \int^{t_1}_{t'+N \tau_0 } dt_1[g'(t) ]^2 \left(  a e^{-i\omega_0 (t_1-t_0) } - a^\dag\, e^{i \omega_0 (t_1-t_0) }  \right)
\left( a e^{-i\omega_0 (t_2-t_0) } - a^\dag e^{i \omega_0 (t_2-t_0) }  \right) \nn
\approx & 1 - i \frac{1 }{ \omega_0}g'(t) \left(  a e^{i \omega_0 t_0}(e^{-i\omega_0 t}-e^{-i \omega_0 t'} ) + a^\dag e^{-i \omega_0 t_0} (e^{i\omega_0 t}-e^{i \omega_0 t'} )\right) \nn
       &- \frac{1 }{ 2}\, \left(\frac{g'(t)}{ \omega_0} \right)^2 \Bigg\{ \left( a e^{i \omega_0 t_0} (e^{-i\omega_0 t}-e^{-i \omega_0 t'} ) +   a^\dag \, e^{-i \omega_0 t_0}  (e^{i\omega_0 t}-e^{i \omega_0 t'} )\right)^2 \nn
       &-  \left[ 2 i \omega_0 (t-t'-N\tau_0 ) + e^{i \omega_0(t-t')} - e^{-i \omega_0(t-t')} \right] \Bigg\} \nn
\equiv & 1 + {\cal S}^{(1)} (t,t') + {\cal S}^{(2)} (t,t').
\label{eq:Lang_Firsov_interaction_propagator_2}
\end{align}
\end{widetext}
Without an explicit knowledge of $g'(t_1)$ one will not be able to complete the integral appearing in the exponent.
\section{Corrections for lesser Green's functions in time-dependent Lang-Firsov model}
\label{appendix b}
The first- and second-order corrections for lesser Green's function given in Eq. (\ref{eq:total lesser Green's function}) are explicitly shown in this Appendix. At first order of ${\cal S}$, the lesser Green's function reads
\begin{widetext}
\begin{align}
G^{<,(1)} (t,t') &= -i e^{-\int^t_{t'} dt_1 \left( \bar{\varepsilon}(t_1) - \frac{|g'(t_1)|^2 }{ \omega_0}\right)}
\Bigl[\langle \alpha | {\cal S}^{(1)}(t_0,t') X^\dag(t') X(t) |\alpha \rangle +\langle \alpha | X^\dag(t') X(t){\cal S}^{(1)}(t,t_0) |\alpha \rangle  \Bigr] \nn
& =  \frac{i }{ \omega_0} G^{<,(0)} (t,t') \nn
&  \times \Bigg\{  g' (t_0) \left( e^{i \omega_0 t'} - e^{i \omega_0 t_0} \right) e^{-i \omega_0 t_0}\alpha^*
+g' (t_0) \left(e^{-i \omega_0 t'} - e^{-i \omega_0 t_0}  \right) e^{i \omega_0 t_0} \Bigl[ \alpha   + g(t') e ^{i \omega_0 t'} - g(t)  e^{i \omega_0 t } \Bigr] \nn
&  -g' (t)\left( e^{ -i \omega_0 t} - e^{-i \omega_0 t_0} \right)e^{i \omega_0 t_0} \alpha
-g' (t)\left( e^{i \omega_0 t} - e^{i \omega_0 t_0 } \right)e^{-i \omega_0 t_0}\left[ \alpha^* + g(t) e^{-i \omega_0 t}  - g(t')   e^{-i \omega_0 t' }  \right] \Bigg\}.
\label{eq:first ordered lesser Green's function}\end{align}
At second order of ${\cal S}$,
\begin{align}
G^{<,(2)}(t,t')
=&  -i e^{-i \int^t_{t'} dt_1 \bar{\varepsilon}(t_1) } \nn
& \times \Bigl\{
\langle \alpha | {\cal S}^{(2)} (t_0,t') X^\dag(t') X(t) |\alpha \rangle
+\langle \alpha |  X^\dag(t') X(t){\cal S}^{(2)}(t,t_0) |\alpha \rangle
+\langle \alpha | {\cal S}^{(1)}(t_0,t') X^\dag(t') X(t){\cal S}^{(1)}(t,t_0) |\alpha \rangle
\Bigr\} \nn
=&  G^{<,(2,1)}(t,t')+G^{<,(2,2)}(t,t')+G^{<,(2,3)}(t,t'),
\label{eq:second ordered lesser Green's function}
\end{align}
where
\begin{align}
 G^{<,(2,1)}(t,t') =&   -\frac{1 }{ 2 \omega_0 ^2 } G^{<,(0)}(t,t') \Bigl\{ \left( g'(t_0)\right)^2 \left( e^{i \omega_0 t_0}- e^{i \omega_0 t'} \right)^2  e^{-2 i \omega_0 t_0} \left( \alpha^*\right)^2 \nn
& + \left[ g'(t_0) \right]^2 \left( e^{-i \omega_0 t_0} - e^{-i \omega_0 t'} \right)^2 e^{2 i \omega_0 t_0}
\left(\alpha + g(t') e^{i \omega_0 t'} -g(t) e^{i \omega_0 t} \right)^2 \nn
&+ [g'(t_0)]^2 \left(e^{-i \omega_0 t_0} - e^{-i \omega_0 t'} \right) \left( e^{i \omega_0t_0} - e^{i \omega_0 t'}  \right)\Bigl[ 1 + 2 \alpha^* \,
\bigl(\alpha+ g(t')e^{i \omega_0 t'} - g(t) e^{i \omega_0 t} \bigr)\Bigr]\nn
&  - [g'(t_0)]^2 \Bigl[2 i \omega_0 \left( t_0-t' - N^{(t_0,t')} \tau\right) + e^{i \omega_0 (t_0-t')} -e^{-i \omega_0(t_0-t')}  \Bigr]\Bigr\}, \nonumber
\end{align}
\begin{align}
G^{<,(2,2)}(t,t')  =&  -\frac{1 }{ \omega_0^2 } G^{<,(0)}(t,t') \Bigl\{  g'(t_0) g'(t) \left(e^{i \omega_0 t_0} - e^{i \omega_0 t'}  \right)
\left( e^{- i \omega_0 t} - e^{-i \omega_0 t_0} \right) |\alpha|^2 \nn
&  + g'(t_0)g'(t) \left( e^{i \omega_0 t_0} - e^{i \omega_0 t'}\right) \left( e^{i \omega_0 t}- e^{i \omega_0 t_0}  \right)  e^{-2 i \omega_0 t_0} \alpha^*
\left( \alpha^* + g(t) e^{-i \omega_0 t} - g(t') e^{-i \omega_0 t'} \right) \nn
&  + g'(t_0) g'(t)  \left( e^{-i \omega_0 t_0} - e^{-i \omega_0 t'}\right) \left( e^{-i \omega_0 t}- e^{-i \omega_0 t_0}  \right)
e^{2 i \omega_0 t_0}\alpha \left( \alpha+ g(t') e^{i \omega_0 t'} - g(t) e^{i \omega_0 t} \right) \nn
& +g'(t_0) g'(t)  \left( e^{-i \omega_0 t_0} - e^{-i \omega_0 t'}\right) \left( e^{i \omega_0 t}- e^{i \omega_0 t_0}  \right)
\Bigl[\left( \alpha^*+ g(t) e^{-i \omega_0 t} - g(t')e^{-i \omega_0 t'}\right) \nn
&~~~\times \left( \alpha + g(t') e^{i \omega_0 t'} - g(t)  e^{i \omega_0 t} \right) + 1  \Bigr]
\Bigr\}, \nonumber
\end{align}
\begin{align}
G^{<,(2,3)} (t,t')  =&  -\frac{ 1 }{\omega_0 ^2 } G^{<,(0)} (t,t') \Bigl\{ [ g'(t)]^2 \left( e^{-i \omega_0 t} - e^{-i \omega_0 t_0} \right)^2 e^{2 i \omega_0 t_0}  \alpha^2 \nn
&  +  [g'(t)]^2  \left(  e^{ i \omega_0 t} - e^{i \omega_0t_0} \right)^2 e^{- 2 i \omega_0 t_0} \left( \alpha^* + g(t)  e^{-i \omega_0 t} - g(t') e ^{-i \omega_0 t'} \right)^2 \nn
&  + [g'(t) ]^2 \left( e^{-i \omega_0 t} - e^{-i \omega_0 t_0 } \right) \left( e^{i \omega_0 t} - e^ { i \omega_0 t_0 } \right)
\left[1+ 2\alpha \left(\alpha^* + g(t) e^{-i \omega_0 t} - g(t') e^{-i \omega_0 t'}  \right) \right] \nn
&  - [g'(t)]^2 \left[2 i \omega_0 \left( t-t_0 -N^{(t,t_0) } \tau\right) + e^{i \omega_0 (t-t_0)} - e^{- i \omega_0 (t-t_0) }  \right]
\Bigr\}. \nonumber
\end{align}
\end{widetext}
Here, $N^{(t,t')} $ is defined as the quotient of dividing $t-t'$ with $\tau = 2 \pi/ \omega_0$.
\bibliography{time-dependent-Lang-Firsov_reference}

\end{document}